# The Quantitative Relations between Stock Prices and Quantities of Tradable Stock Shares and Its Applications

**Chengling Gou**

Physics Department, Beijing University of Aeronautics and Astronautics

No.37, Xueyuan Road, Heidian District, Beijing, 100083, China

Physics Department, University of Oxford

Clarendon Laboratory, Parks Road, Oxford, OX1 3PU, UK

c.gou1@physics.ox.ac.uk    gouchengling@hotmail.com

**Abstract:** This paper analyzes the quantitative relations between stock prices and quantities of tradable stock shares in Chinese stock markets at six time points by means of Exploratory Data Analysis (EDA) method. It is found the resulting formulae have the same structure but different parameters. This paper also uses these relationships in order to analyse the feasibility of policies for Chinese Government to sell the state-owned shares in Chinese stock markets

# 1  Introduction:

Chinese stock markets (i.e. Shanghai stock market and Shenzhen stock market) are immature and have their own characteristics. First, their structures are different from the most western stock markets. They have two different kinds of stock shares; one is tradable in both stock markets, which are held by ordinary investors, and the other is not tradable in both stock market which are held by the government known as state-owned shares  Zhou  2003  . Second, the prices of stock shares are highly influenced by the quantity of their tradable stock shares as shown in figure 1. Moreover, Chinese government always intends to sell state-owned shares in Chinese stock markets so that this causes great potential system risk to the markets. For example, Chinese government wanted to sell the state-owned shares in Chinese stock markets from 12 June 2001 to 24 June 2002. This action made the markets crash down. Shanghai Index fell from 2245 to 1335. Eventually the government had to stop their selling action.

Why did the government fail to sell their shares? Is there a win-win solution? More than 2000 proposals about selling state-owned shares were handed in to Chinese Securities Regulation Commission after it called for proposals about selling state-owned shares in Nov. 2001. Among these proposals, the consensus of markets prefers the share-adjusted plan, which means to adjust the proportion of tradable shares to state-owned shares. This plan has two operation proposals: one is to shrink the state-owned shares, which is called the shrinking totally tradable plan; the other is to split-up tradable shares, which is called the split-up totally tradable plan. Specialists and ordinary investors think these two plans are equivalent. Is this really true?

This paper first analyzes the quantitative relations between stock prices and quantities of tradable stock shares in Chinese stock markets at six time points by means of Exploratory Data Analysis (EDA) method. Then this paper uses these relations to analyse the feasibility of policies for Chinese Government to sell the state-owned shares in Chinese stock markets

## 2  The Quantitative Relations between Stock Prices and Quantities of Tradable Stock Shares

From both Shanghai stock market and Shenzhen stock market, I select 240 stocks which belong to five groups: medicine manufacture, steel manufacture, appliance manufacture, car manufacture and gasoline manufacture. The criterion of selecting sample groups is that the groups have wide span of quantities of stock shares. I analyze the relations between stock prices and quantities of tradable stock shares at six time points by using the 60-day-average prices. By means of Exploratory Data Analysis (EDA) method (Hoaglin, 1998,), I obtain the quantitative relations between stock prices and quantities of tradable stock shares at six time points (referring to figure 2, Gou, 2003), which are shown as following.

Make $x_{0i}$ represent the quantity of tradable stock shares, $y_{0i}$ represents the 60-day-average prices of stocks and $\hat{y}_i$ represent the formula value of each transformed $y_i$. By making power transformation as (Hoaglin, 1998,):

$$x_i = \ln x_{0i}, \quad y_i = \frac{y_{0i}^{-0.1} - 1}{-0.1},$$

six formulas are obtained as:

$$\hat{y}_1 = 2.279 - 0.2968(x_1 - 18.42), \qquad 2001/6/29 \qquad (1)$$

$$\hat{y}_2 = 2.059 - 0.3015(x_2 - 18.42), \qquad 2001/12/31 \qquad (2)$$

$$\hat{y}_3 = 1.998 - 0.3011(x_3 - 18.42), \qquad 2002/6/30 \qquad (3)$$

$$\hat{y}_4 = 1.855 - 0.2733(x_4 - 18.60), \qquad 2002/12/27 \qquad (4)$$

$$\hat{y}_5 = 1.883 - 0.1929(x_5 - 18.60), \qquad 2003/7/1 \qquad (5)$$

$$\hat{y}_6 = 1.700 - 0.1257(x_6 - 18.60), \qquad 2003/12/31 \qquad (6)$$

The statistic rule can be concluded as

$$\hat{y}_i = a_i - b_i(x_i - c_i) \qquad (7)$$

where *a* is associated with the average prices of the whole stock markets; *b* represents the price structure according to the quantities of stock shares; and *c* is determined by the size of quantities of stock shares in the whole markets.

We can find out that because the average prices reduced from Jane, 2001 to Jane, 2002, parameter *a* became smaller but parameter *b* stayed almost the same, seeing figure 3 and figure 4. After Jane, 2002, parameter *a* still became smaller and parameter *b* became larger gradually. This change is probably due to the new concept about value investment proposed by institutional investors. However, the structure of formula (7) still holds all the time.

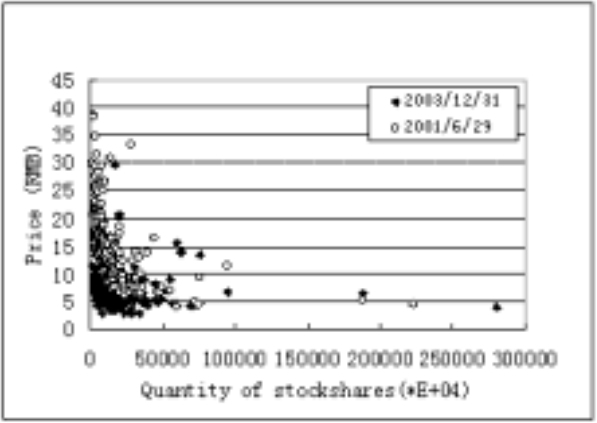

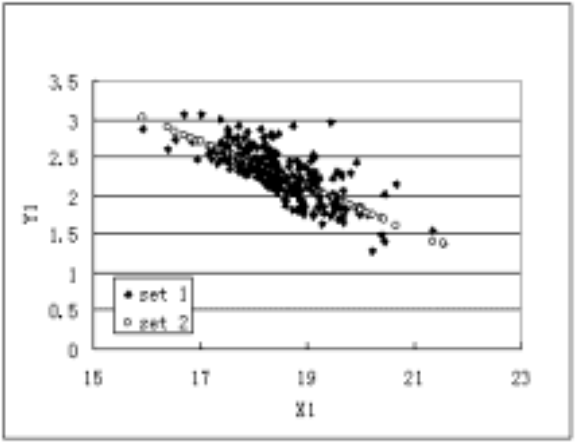

**Figure1** Showing the Relationships between Prices of Stock Shares and Their Quantities.

**Figure2** Showing Power Transformed Prices $y_1$ and Quantities $x_1$ with Set 1 and Their Simulated Values with Set 2.

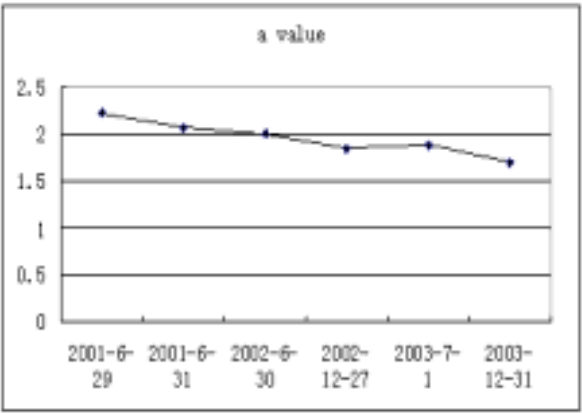

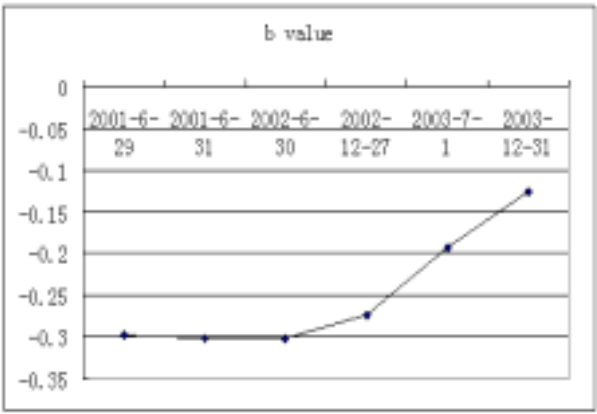

**Figure 3** Showing the *a* Values at the Six Time Points

**Figure 4** Showing the *b* Values at the Six Time Points

## 3  Application

### 3.1 The Reason for the Failure of Selling State-owned Shares

Although Chinese government only wanted to sell 10% state-owned shares accompanying with IPO and refunding of listing companies during the period from 12 June 2001 to 24 June 2002, the markets had the anticipation that the state-owned shares would be totally tradable in the markets in the uncertain future. This made the markets crash. According to formula (7), we can understand why the markets crashed. I use formulas (1) and (3) to calculate the prices of sample companies under the condition that the state-owned shares would be totally tradable. This is to say, use total quantity of shares (tradable shares + state-owned shares) in formulas (1) and (3) to calculate the simulated prices $\tilde{y}_{01}$ and $\tilde{y}_{03}$ which I call as totally-tradable prices. Fig.5 shows that $\tilde{y}_{01}$, $\tilde{y}_{03}$ are much lower than $y_{01}$ $y_{03}$ which are 60-day average prices on 2001/6/29 and 2002/6/30, respectively. Max, median, min and average of $y_{01}$, $y_{03}$, $\tilde{y}_{01}$, $\tilde{y}_{03}$ and the differences of these prices are listed in table 1. From table 1, we can see the average prices would drop at least 34% if the state-owned shares would be totally tradable at that time. Moreover, stock markets will crash once they form falling tendency because stock markets are positive feedback systems. In fact, the Chinese stock markets would drop further if the government had not stopped selling state-owned shares.

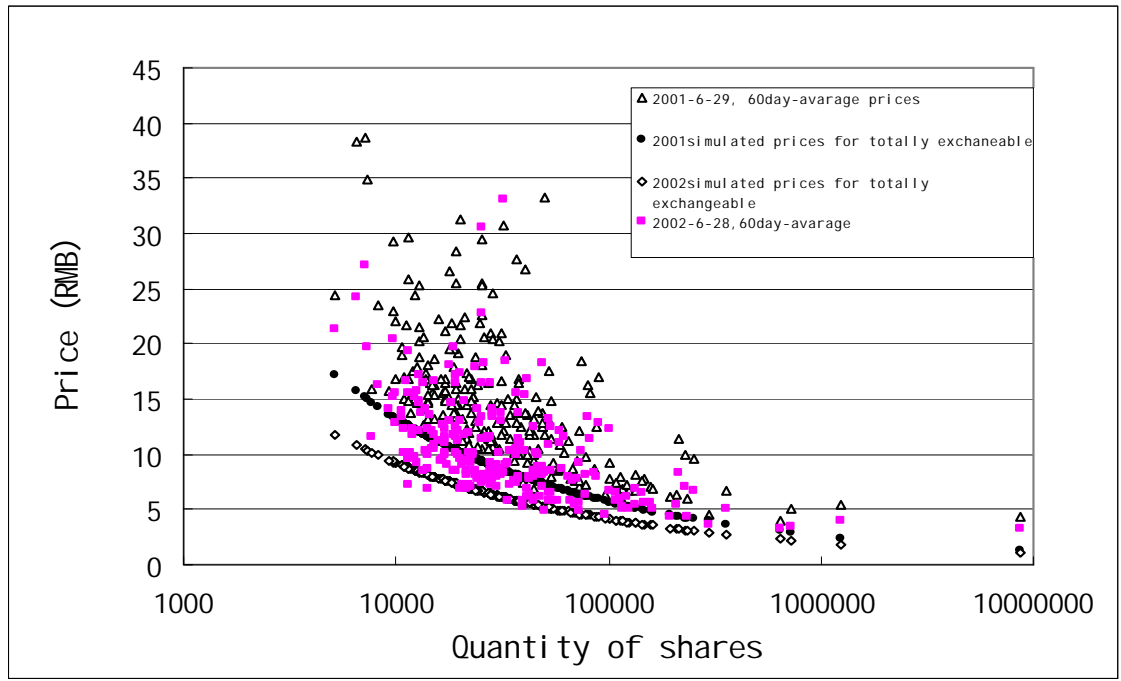

**Figure 5** Real Market Prices and Simulated Totally-tradable Prices

**Table 1** Real Market Prices and Totally-tradable Prices and Their Differences

|  | max | median | min | average |
| --- | --- | --- | --- | --- |
| $y_{01}$ (60-day average real market prices) | 38.637 | 13.759 | 3.967 | 14.638 |
| $\tilde{y}_{01}$ ( simulated prices of formula (1)) | 17.214 | 8.974 | 1.310 | 8.956 |
| $(y_{01} - \tilde{y}_{01})/y_{01} \times 100$ (%) | 78 | 32 | -9 | 34 |
| $y_{03}$ (60-day average real market prices) | 33.105 | 9.907 | 3.191 | 8.965 |
| $\tilde{y}_{03}$ ( simulated prices of formula (3)) | 11.971 | 6.391 | 1.033 | 6.362 |
| $(y_{03} - \tilde{y}_{03})/y_{03} \times 100$ (%) | 82 | 35 | -20 | 34 |

**3.2 Effect on the Markets of the shrinking Totally Tradable Plan**

Zhang proposed the formula for the shrinking totally tradable plan (Zhang, 2002):

Shrinking proportion = (real market prices)/ (net asset per share)     (8)

New total shares = (tradable shares) + (state-owned shares)/ (shrinking proportion)   (9)

To calculate the simulated prices $\tilde{y}_{11}$ under the condition of shrinking totally tradable plan on 2001/6 /29, I use formulas(1), (8) and(9), in which tradable shares, state-owned shares and net assets per share are obtained from the half-year reports of listing companies in June 2001 and $y_{01}$ are 60-day average real market prices on 2001/6 /29 . As shown in Fig. 6  most $\tilde{y}_{11}$ are lower than $y_{01}$, which means the shrinking totally tradable plan would make prices of stock shares fall if it were carried out at that time. Table 2 shows prices would fall at 14% averagely according to this calculation.

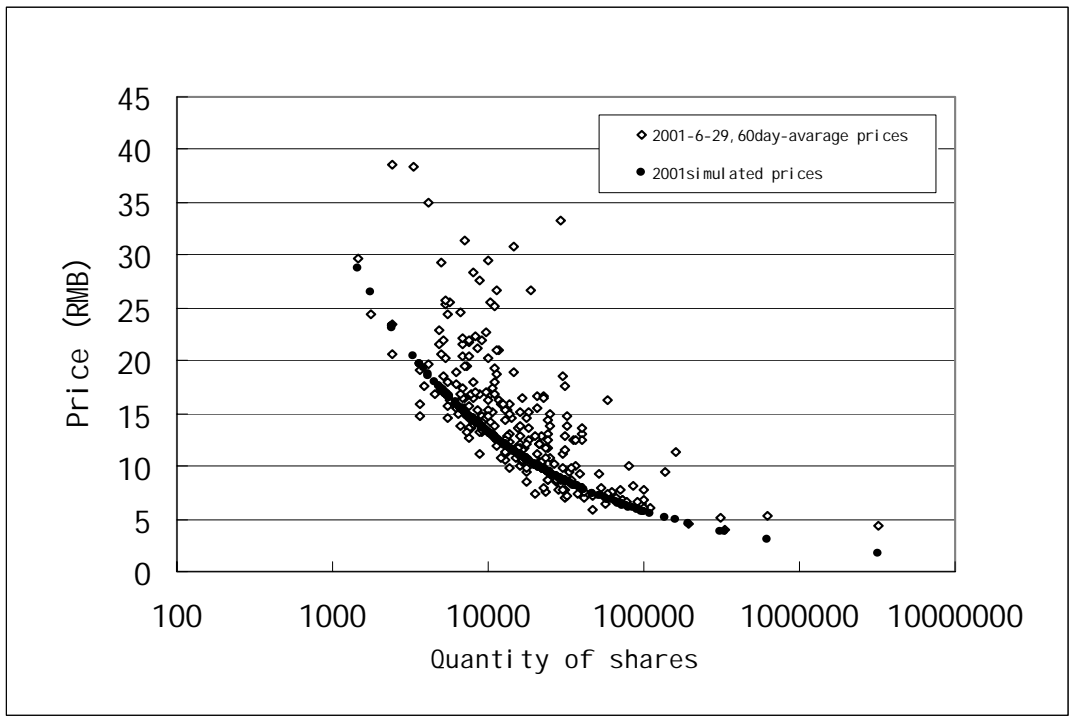

**Figure 6** Real Market Prices and Simulated Prices of Shrinking Totally-Tradable Prices

**Table 2** Real Market Prices and Shrinking Totally-Tradable Prices and Their Differences

|  | max | median | min | average |
|---|---|---|---|---|
| $y_{01}$ (60-day average real market prices) | 38.637 | 13.759 | 3.967 | 14.638 |
| $\tilde{y}_{11}$ ( simulated prices of formula (1)) | 28.658 | 11.549 | 3.041 | 11.874 |
| $(y_{01} - \tilde{y}_{11}) / y_{01} \times 100$ (%) | 73 | 12 | 36 | 14 |

**3.3 Effect on the Markets of the Split-up Totally Tradable Plan**

Zhang gave the following formulas for the split-up totally tradable plan (Zhang 2002):

Split-up proportion = (real market prices)/ (net asset per share)　　　　(10)

New total shares = (state-owned shares) + (tradable shares)*(split-up proportion)　　　(11)

Split-up prices = (real market prices)/ (split-up proportion)　　　　(12)

I use formulas(10),(11), (12) and (1) to calculate the simulated prices $\tilde{y}_{21}$ under the condition of the split-up totally tradable plan on 2001/6 /29, in which tradable shares, state-owned shares and net assets per share are obtained from the half-year reports of listing companies in June 2001 and 60-day average real market prices $y_{01}$ on 2001/6 /29 . As shown in Fig. 7 nearly all $\tilde{y}_{21}$ are greater than split-up prices $\bar{y}_{01}$, which means the split-up totally tradable plan would make prices of stock shares go up potentially if it were carried out at that time. Table 3 shows prices would go up at 159% averagely according to this calculation.

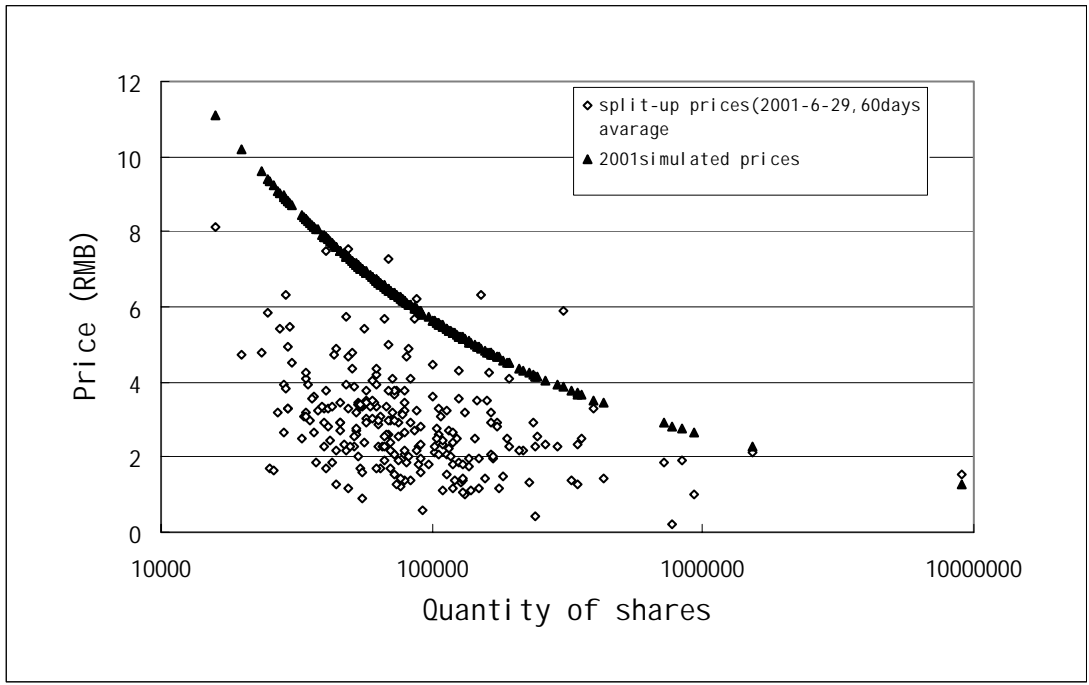

**Figure 7** Split-up Prices and Simulated Prices of Split-up Totally-Tradable Prices

**Table 3** Split-up Prices and Split-up Totally-tradable Prices and Their Differences

|  | max | median | min | average |
| --- | --- | --- | --- | --- |
| $\bar{y}_{01}$ (split-up prices) | 8.131 | 2.760 | 0.207 | 2.916 |
| $\tilde{y}_{21}$ (simulated prices of formula (1)) | 11.123 | 6.357 | 2.257 | 6.320 |
| $(\tilde{y}_{21} - \bar{y}_{01}) / \bar{y}_{01} \times 100$ (%) | 1262 | 126 | 35 | 159 |

**3.4 Suggestions to Policy-makers and Foreign Investors**

In Chinese stock markets, the quantitative relations between stock prices and quantities of tradable stock shares are the key factor to determine stock prices. Because of unawareness of this, Chinese government failed to sell state-owned shares in the markets. The shrinking totally tradable plan will make the markets form a going-down anticipation, which will form bear markets. So the shrinking totally tradable plan is not doable. In contrast, the split-up totally tradable plan could be a win-win solution for Chinese government to sell state-owned shares in Chinese stock markets. Because this plan will make the markets form going-up anticipation, which will probably make money flow into the markets so as to produce money-making effect. This will attract more money into the markets so that bull markets could form. The government could sell state-owned share in bull markets successfully.

In order to make the split-up totally tradable plan carried out successfully, two steps should be followed: first, split-up the tradable shares so that the state-owned shares obtain the right to be traded in the markets; second, make a schedule for state-owned shares to be traded in the markets, that is to say, the state-owned share can not be sold in the markets at the same time, but must be sold gradually. If the state-owned shares which are huge flood into the markets, the markets can not afford.

Chinese stock markets are opened to foreign investors under certain condition. Foreign investors can invest in Chinese stock markets through Qualified Foreign Investment Institutions (QFII). Before they make investment decision, they need to know the characteristics of Chinese stock markets. First, they must be aware of the quantitative relations between stock prices and quantities of tradable stock shares; second, they must pay attention to the system risk of Chinese stock markets due to the state-owned shares. The proportion of state-owned shares to total shares in the Chinese stock markets is two thirds. Therefore, Chinese government is a huge player in the markets actually. It is potential system risk that whatever decisions the government will make about state-owned shares.

**4　Conclusion:**

1) In Chinese stock markets, the quantitative relations between stock prices and quantities of tradable stock shares are the key factor to determine stock prices. All participants in Chinese stock markets including investors and Chinese government need to pay attention to it when they make any decisions.

2) The split-up totally tradable plan could be a win-win solution for Chinese government to sell state-owned shares in Chinese stock markets.

**Acknowledgements**

This research is supported by Chinese Overseas Study Fund. Thanks Mr. Pingsheng Qian for preparing data. And thanks Professor Neil Johnson for his suggestion about plotting *a* and *b* values.